\documentclass[a4paper,11pt]{article}
\pdfoutput=1 

\usepackage{jcappub} 

\usepackage[T1]{fontenc}
\usepackage{times}
\usepackage{graphicx,latexsym, amssymb, amscd, psfrag}
\usepackage{color}
\usepackage{morefloats,rotating,float}
\usepackage{gensymb}
\usepackage{multirow,array}
\usepackage{adjustbox}
\usepackage{gensymb}
\usepackage{mathtools}
\usepackage{framed, caption}
\usepackage{subfigure}
\usepackage{hyperref,breakurl}

\title{An {\it AstroSat}/UVIT study of galaxies in the cluster \abell}

\author[a,1]{Smriti Mahajan\note{Corresponding author}}
\author[a]{Kulinder Pal Singh} 
\author[b,c]{Somak Raychaudhury} 
\affiliation[a]{Department of Physical Sciences, Indian Institute for Science Education and Research Mohali- IISERM, Knowledge City, Manauli,  Punjab 140306, India }%
\affiliation[b]{Department of Physics, Ashoka University, Sonipat, Haryana 131029, India}%
\affiliation[c]{Inter-University Centre for Astronomy and Astrophysics, Pune, Maharashtra 411007, India}%

\emailAdd{mahajan.smriti@gmail.com}
\emailAdd{kps@iisermohali.ac.in}
\emailAdd{somakrc@gmail.com}

\def\fuv{{\it FUV }}
\def\nuv{{\it NUV }}

\def\arcsec{{^{\prime\prime}}}
\def\arcmin{{^{\prime}}}
\def\g{{\it Galex }}
 
\definecolor{grey}{rgb}{0.5,0.6,0.7}

\definecolor{amber}{rgb}{1.0,0.49,0.0}

\newcommand{\abell}{{Abell~2199}}
\newcommand{\mcg}{{Z224-44}}

\usepackage{aas_macros}
\usepackage{hyperref} 
\hypersetup{colorlinks,citecolor=blue,linkcolor=blue,urlcolor=blue}

 \abstract{We present the newly acquired data for an {\it AstroSat}/UVIT field centered on a face-on spiral starburst galaxy UGC~10420, located in the X-ray bright cluster Abell 2199 ($z = 0.031$). 
 We have analysed the \fuv BaF$_2$ data for this field along with the archival \fuv and \nuv data from the {\it Galex} mission, optical photometric data from the SDSS, spectroscopic data 
 from the literature, 
 and low-frequency radio data from the LoTSS survey, respectively. The stars were separated from the galaxies using the SDSS photometric pipeline classification, while the spectroscopic 
 redshifts available for 35\% of the detected UVIT sources were used to identify member galaxies of the cluster \abell. We find that (a) the non-cluster galaxies are on average fainter than the 
 cluster galaxies at fixed magnitude, (b) stars and galaxies are indistinguishable in the $r$ vs $NUV-r$ plane, and (c) bright stars are $\sim 1.5$ mag bluer than the galaxies in the 
 $FUV-r$ vs $NUV-r$ colour-colour plane. Besides UGC~10420 which is the only known cluster galaxy with an extended-UV disk, we identify five more galaxies with asymmetric \fuv 
 morphology and extended radio emission in this field. All the asymmetric member galaxies of \abell, lie within the virial boundaries of the cluster.
 This observation, together with the fact that these asymmetric cluster galaxies have low-frequency radio tails or \fuv emission pointing away from the cluster centre leads us to hypothesise that
 these galaxies are likely undergoing ram-pressure stripping (RPS) under the influence of cluster-environment related mechanisms. A comparison of optical and \fuv star formation 
 rate of UVIT detected galaxies shows enhanced star formation in half of the RPS candidates, suggesting that environment-related mechanisms may lead to a burst of star formation 
 in RPS galaxies.  

 Our analysis indicates the presence of at least two more groups or clusters at $z \sim 0.077$ and $0.260$, coincident with \abell~along the line of sight of the field of view studied here.      }

\begin{document}
 \maketitle
\flushbottom

 \section{Introduction}
 \label{intro}

 Galaxy clusters are the largest gravitationally bound objects in the Universe. They form at the intersection of large-scale filaments, therefore allowing for sampling of 
 a wide range of properties of galaxies such as colours, luminosity, stellar mass, star formation rate (SFR), gas and dust properties etc. This feature makes clusters of galaxies 
 ideal laboratories to study the evolution of galaxies over a continuous range of `environments', quantified either by the number density of galaxies, or the density of the intra-cluster 
 medium (ICM). 
 
 The ultraviolet imaging telescope \citep[UVIT;][]{tandon17, tandon20} aboard the multiwave Indian satellite mission {\it AstroSat} \citep{singh14}, with its wide field of view of 
 $\sim 28\arcmin$ is the perfect choice for imaging 
 clusters in the far ultraviolet ({\it FUV}) waveband. The UVIT offers the best spatial resolution of $1.3\arcsec-1.5\arcsec$ \citep{tandon20}, relative to similar missions, 
 e.g. Galaxy evolution explorer ({\it Galex}), Swift-UVOT and XMM-OM, observing in the \fuv waveband. The unprecedented resolution and large field of view together make UVIT a great 
 tool for exploring the recent star formation history of galaxies in and around clusters.
 
 In this era of wide field surveys such as the two-degree Field Galaxy Redshift Survey \citep[2dFGRS;][]{colless99} and the Sloan Digital Sky Survey 
 \citep[SDSS;][]{abazajian03}, the regions in the outskirts of clusters have emerged as fields of great interest for studying the evolution of galaxies. These regions offer three-fold advantage
 for studying the evolution of galaxies: 
 \begin{itemize}
 \item The intermediate density environment is ideal for gravitational interactions among galaxies comoving in the same velocity field under 
 the gravitational influence of the nearest massive cluster \citep[e.g.][]{gonzalez09, mahajan12}. 
 \item The density, and therefore the influence of ICM is expected to be weak on galaxies beyond the virial radius (1-2$~R_{200}$). This offers an opportunity to observe galaxies 
 undergoing accelerated evolution in the absence of cluster-specific environmental mechanisms. 
 \item These regions are bound by very dense cluster on one side and low density field on the other, therefore offering a smooth transition in the parameter(s) used to quantify environment.  
 \end{itemize}

 The galaxies at the periphery of clusters experience a unique environment. While the influence of the ICM may still be feeble at these distances from the 
 centre of the cluster, the unidirectional velocity fields pulling the galaxy towards the centre of the cluster are strong \citep{gonzalez09}. The increased density of galaxies on the periphery 
 of the cluster due to the culmination of large-scale filaments approaching from different directions provides better opportunities for galaxies to interact gravitationally. Assuming that most of the 
 galaxies being assembled are young and gas-rich, this scenario may trigger an episode of intense star formation in a small clustercentric region on the boundary of the cluster 
 \citep[e.g.][]{mahajan12}, consequently leading to exhaustion of gas in galaxies even before they enter the cluster. This is an underrated, but important way in which galaxies can be 
 pre-processed prior to entering the clusters \citep[e.g.][]{porter05, mahajan12, kuchner22, salerno22, ruiz23}.
 
 The outskirts of clusters also provide an insight into the evolution of unique population of galaxies which have crossed the cluster centre at least once, yet have not had the time to virialize. 
 These galaxies, mostly found between 1--2 $r_{virial}$, are called `backsplash' galaxies \citep{gill05}. The infalling and backsplash galaxies are not only found at similar distance from the 
 centre of the cluster, but may also be alike in morphology, colour etc.. In the observed data however, the backsplash galaxies can be differentiated from the 
 infalling ones based on their line-of-sight velocity, and by analysing the impact of the cluster environment on their observed properties \citep{mahajan2011, ruiz23}.   
  
 Literature based on observations 
 suggests that backsplash galaxies are older, have lower star formation rate and higher fraction of quiescent galaxies relative to the new entrants to the cluster 
 \citep{mahajan2011, kuchner22, ferreras23}. In fact, \cite{mahajan2011} used optical data for more than 260 galaxy clusters from the SDSS, together with dark matter simulations to show 
 that the star formation in a galaxy is likely to be completely quenched in a single passage through the cluster centre. Even the ultraviolet SFR of galaxies
 on the outskirts of the Coma cluster ($\sim 100 h^{-1}$ Mpc) is found to be in between that of the field and the cluster interiors \citep{mahajan2018}. These observations highlight the importance of 
 studying the transitional populations of galaxies on the cluster outskirts to further our understanding of the evolution of galaxies, and assembly of the cosmic-web.

 With this in mind, we initiated a campaign to explore the outskirts of nearby clusters in the \fuv waveband. We centred the $28\arcmin$ UVIT field of view
 on galaxies known to have higher than average SFR (as estimated using the SDSS spectroscopic data) in \cite{mahajan12}. Aided by ancillary data in other wavebands, 
 these newly acquired {\it AstroSat}/UVIT data are furthering our understanding of the properties and evolution of the targeted galaxies \citep{mahajan23}. These data have also provided an 
 opportunity to study the properties of a variety of galactic and extragalactic sources detected by the UVIT \citep{mahajan22}, thereby creating a benchmark for other similar studies.  
 
 In this paper, we study a field in the cluster \abell~($z = 0.031$). \abell~is an X-ray bright cluster, with a massive radio-loud galaxy NGC~6166 at its core. Together with its neighbouring  
 cluster Abell~2197 ($z=0.030$), and several other groups of galaxies, \abell~forms a large supercluster \citep{rines01, rines02}. A field centred around $0.45$ r$_{virial}$ of this dynamic cluster 
 provides a golden opportunity to study plethora of interesting galaxies, many of which are in a transition phase. 
 Throughout this work we use concordance $\Lambda$ cold dark matter cosmological model with $H_0 = 70$ km s$^{-1}$ Mpc$^{-1}$, $\Omega_\Lambda = 0.7$ 
 and $\Omega_m = 0.3$ to calculate distances and magnitudes.

 \begin{figure}
 \centering{
 {\includegraphics[scale=0.5]{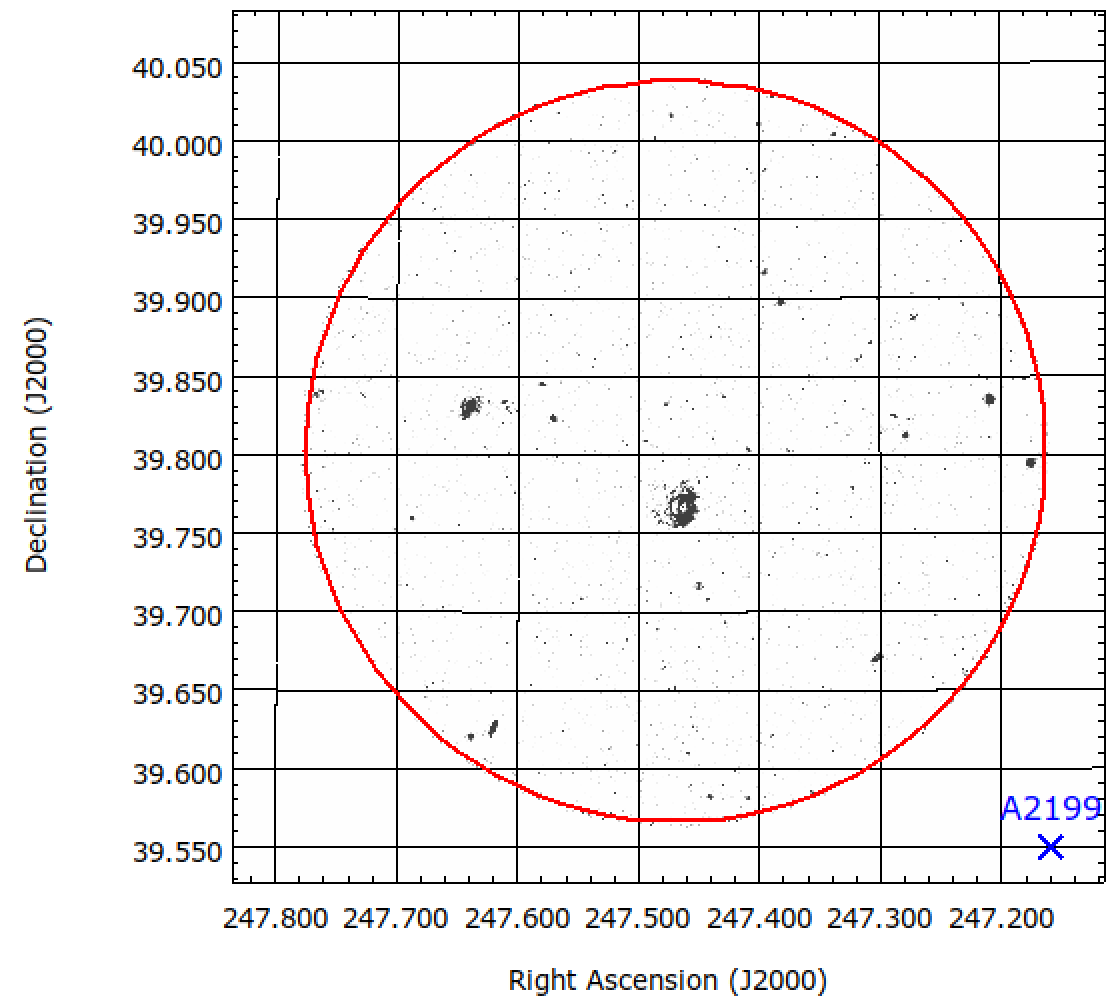}}}
 \caption{The UVIT field of view with the {\it red} circular region centred at $\alpha = 247.468^{\circ}$ and $\delta = +39.802^{\circ}$, having a radius of $14.15\arcmin$ analysed in 
 this work. The blue `X' symbol shows the x-ray centre of Abell~2199 \citep{rines02}. The image is orientated such that north is up and east is on the left.  } 
 \label{fov}
 \end{figure}

\section{Data}
\label{data}

\subsection{Ultraviolet imaging and data processing}
\label{uvdata}
 
 We observed a field centred on the spiral galaxy UGC~10420 \citep[$\alpha = 247.463^{\circ}~{\rm and}~\delta = 39.766^{\circ}$;][]{mahajan23}, in \abell~($z = 0.031$) 
 with the {\it AstroSat}/UVIT \citep{kumar12, singh14, tandon17}. 
 This field (observation ID: A05\_063T03\_9000003066; PI: Smriti Mahajan; observation date: 30-July-2019) was observed by the UVIT simultaneously in the broadband 
 BaF$_2$ \fuv filter \citep[mean $\lambda = 1541~\AA$; $\Delta\lambda = 380~\AA$;][]{tandon17, tandon20}, and visible wavebands ($3040-5500~\AA$). The latter data however,
 are only used for data reduction. The raw data from the mission are processed by the Indian Space Science Data Centre (ISSDC), and the Level 1 (L1) data from all instruments are provided to the users.  The L1 data were processed using the CCDLAB  software \citep{postma17}, following the procedure described in \cite{postma21}. A detailed account of the steps followed for the reduction of data are well described in  \cite{mahajan22}, but a very brief description is given below for completeness.
 
The CCDLAB software pipeline takes the L1 zip file as input and performs various corrections on individual images before performing the astrometric correction to coalign and stack them together to create a deep field image. The automatic procedure for optimising the point source function was then initiated to remove the residual drift effect, and yielding a full width half maximum of $< 1.2\arcsec$ for this image. Following this, the world coordinate system (WCS) solution was applied by mapping the pixel coordinates of the sources in the image to the GAIA catalogue using the Astroquery utility \citep{ginsburg19}, through CCDLAB \citep{postma20}. The image was finally derotated to match the sky coordinates, and used for further scientific analysis. The UVIT data for \abell~have a useful exposure time of 9600.3 seconds and resolution $<1.2\arcsec$. The final image is shown in Figure~\ref{fov}, with the UVIT field of view and the centre of \abell~marked for reference.
 
 The \fuv sources were detected using the {\sc sextractor} software \citep[SE henceforth;][]{bertin96}. Sources were detected on a convolved version of the image (Gaussian kernel of 
 FWHM $ = 4.0$ pixels) in order to prevent shredding of larger galaxies. Other de-blending and detection parameters used by {\sc sextractor} were chosen after extensively 
 testing a wide range of values and visually testing the processed image, and are similar to the ones mentioned in  \cite{mahajan22}. The background was detected by 
 {\sc sextractor} in the `AUTO' mode. 
 
 A deeper analysis of the generated catalogue showed that four of the larger galaxies in the field were broken into multiple components. For two of these galaxies 
 (LEDA\footnote{Lyon-Meudon Extragalactic Database (LEDA), \cite{leda}} 58382 and LEDA 2153801), which were split into two components each, the total counts estimated 
 by the SE were summed up to get the total flux. For the other two (UGC 10420 and UGC 10429), we used the SAOImageDS9 software package \citep{ds9} to get the total counts using the 
 method described in \cite{mahajan23}. To do so, we chose an aperture size of $1.5\arcmin$ centred at the optical nucleus for UGC 10420, and of $0.8\arcmin$ for UGC 10429, respectively.

\begin{figure}
 \centering{
 {\includegraphics[scale=0.5]{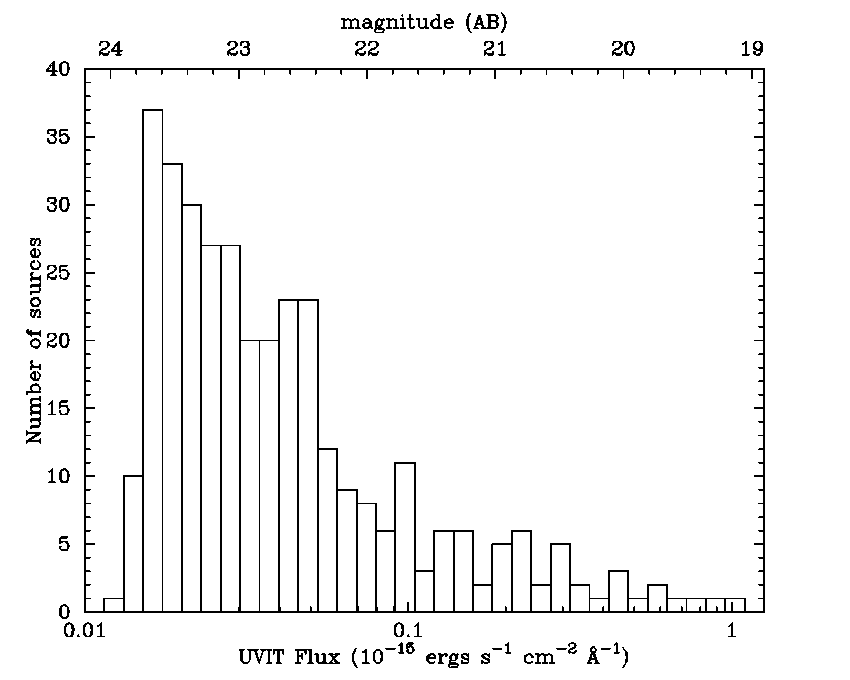}}}
 \caption{The distribution of UVIT \fuv flux corrected for the Milkyway extinction for all the 352 sources analysed in this work. The top axis shows the AB magnitudes for the same.}
 \label{flux}
 \end{figure}

 Following this methodology, we detected 352 unique sources in the UVIT BaF$_2$ filter, which form the primary sample for this paper. The integrated counts for all the sources were then 
 normalised by the total integration time, and converted to fluxes using the conversion factors provided by \cite{tandon20}. These fluxes were corrected for the extinction due to the dust in 
 our galaxy by assuming a Milkyway-like extinction curve \citep{calzetti00}. We assume $A_{FUV} = 8.06 E(B-V)$ \citep{bianchi11}, where the $E(B-V)$ estimates of \cite{schlafly11}
 for all the identified sources were obtained from \url{https://irsa.pac.caltech.edu/applications/DUST/}. The mean reddening excess, $E(B-V) = 0.0087 \pm 0.0012$ for our sample. 
 Figure~\ref{flux} shows the distribution of the extinction corrected fluxes and the corresponding magnitudes for all the detected sources.

 \subsection{Optical data}
 
 We found optical counterparts for 337/352 sources in the Sloan Digital Sky Survey \citep[SDSS, data release 16;][]{sdss16} photometric catalogue. The optical counterparts were found
 by searching for an analogue within $5\arcsec$ of each \fuv source, and having $r < 22.7$ mag\footnote{The median $5\sigma$ depth for SDSS photometric observations is 
 $r = 22.7$ mag.}. More than 95\% of the matches were found within a radius of $1.4\arcsec$. 281 of these sources are identified as galaxies, while the rest of the 56 are stars in the 
 Milkyway according to the SDSS photometric pipeline. It is noteworthy that on visual inspection many of the `stars' appear to be faint distant galaxies, or compact sources 
 on the outskirts of large nearby galaxies \citep[e.g.][]{werk10} in the multi-band SDSS images. A confirmation of the same however, would require higher resolution images and spectroscopic 
 data for these sources, and in this work we continue to consider them as stars. 
   
 An extensive spectroscopic survey of the cluster \abell~has been performed by \cite{song17}. They combined 775 new redshifts from the MMT/Hectospec observations with the existing ones in 
 the literature to construct a large catalogue of 1624 objects in the field of \abell. Their catalogue is spectroscopically 77\% complete to $r_{petro,0} < 20.5$ mag. By searching 
 for an optical counterpart for each of our confirmed \fuv source within a radius of $5\arcsec$, we found spectroscopic redshifts for 123 UVIT sources, more than 96\% of which are matched 
 within $\leq 2.0\arcsec$. Of these, 42 redshifts were taken from \cite{rines08}, 32 from SDSS (DR 12), 48 were added by the MMT/Hectospec observations of 
 \cite{song17}, and one was taken from \cite{huchra12}, respectively. Amongst these 123 sources, 57 are classified as extended sources while 66 are point sources according to the SDSS
 pipeline. A few rows from the complete catalogue of 352 sources detected in this UVIT image are presented in Table~\ref{tab:cat}. The complete catalogue is available online.      
 
   \begin{table*}
 \caption{The object ID, sky coordinates, UVIT flux and the uncertainty in the flux, object type (1: QSO; 3: galaxy; 6: star), $r$-band magnitude and redshift for all the 352 sources detected in 
 the \fuv~image used in this paper. (A complete version of this table is available online here (Supplementary materials)).}
 \begin{center}
 \begin{tabular}{ lccccccc}     
 \hline
    ID  & $\alpha$ & $\delta$  & Flux  ($10^{-15}$ &  $\Delta$(Flux) ($10^{-15}$    & Object  &  $m_r$  & Redshift \\ 
        &    (J2000)    &   (J2000)      &  ergs/s/cm$^2$/\AA)  & ergs/s/cm$^2$/\AA)  & type      &  mag &   $(z)$\\ \hline
 \hline
   1  & 247.443  &   40.034    &   0.0173    &  0.0028   &  3    &  20.82    &  -    \\ 
   2    & 247.389   &  40.028   &    0.0835  &    0.0061    & 3    &  19.25   &   -    \\ 
   3    & 247.542   &  40.028   &    0.0356   &   0.0040   &  3    &  20.78   &   -    \\ 
   4   &  247.416   &  40.023    &   0.0178   &   0.0028   &  3     & 19.91   &   -    \\ 
   5   &  247.461  &   40.022   &    0.0195   &   0.0030    & 3    &   20.56   &   -    \\ 
   6   &  247.473   &  40.016   &    0.5242   &   0.0147   &  3   &   17.46     & 0.048   \\ 
   7   &  247.567    & 40.018    &   0.0960   &   0.0064   &  3    &  19.10     & -    \\ 
   8   &  247.400  &   40.011   &    0.4432   &   0.0137   &  3    &  18.31     & 0.075   \\ 
   9   &  247.590    & 40.013   &    0.0415    &  0.0043  &   3     & 20.66     & -    \\ 
  10   &  247.393   &  40.011    &   0.0244   &   0.0033   &  3    &  19.92     & -    \\ 
\hline
\end{tabular}
 \end{center}
 \label{tab:cat}
 \end{table*}

  \begin{figure}
 \centering{
 {\includegraphics[scale=0.5]{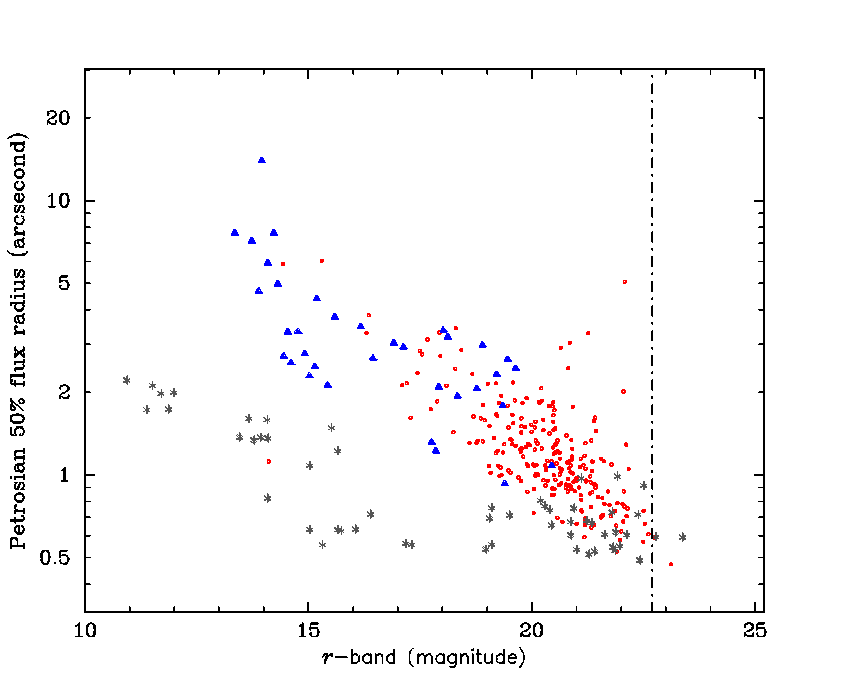}}}
 \caption{The distribution of 50\% Petrosian radius as a function of extinction corrected $r$-band magnitude for the optical counterparts of our UVIT sources. The sources
 are separated into stars {\it (grey asteriks)} and galaxies {\it(red points)} by using the SDSS classification. The member galaxies of \abell~are highlighted as {\it blue triangles}. The 
  {\it dot-dashed line} represents the completeness limit for the SDSS photometric catalogue.}
 \label{rmag-r50}
 \end{figure}

 We identified the member galaxies of \abell~based on the cluster properties derived by \cite{rines16}. These authors found that \abell~($z=0.031$) has a velocity dispersion, 
 $\sigma_v = 676^{+37}_{-32}$ km s$^{-1}$, and mass $M_{200} = (2.39 \pm 0.77) \times 10^{14}$ M$_\odot$. In this paper, we define all galaxies within $\pm3\sigma_v$ of the 
 cluster's mean redshift as the members of the cluster Abell 2199. All other galaxies, with or without redshift are henceforth referred to as the `other' galaxies, although some of them, 
 especially the ones close to the faint-end limit of SDSS (e.g. Figure~\ref{rmag-r50}), might be members of the cluster as well.  

 In Figure~\ref{rmag-r50} we plot the Petrosian radius containing 50\% of the total flux in the $r$-band ($R_{50}$) as a function of the extinction-corrected $r$-band magnitude for all the optical
 counterparts of the UVIT sources sub-classified into stars, cluster galaxies and other galaxies, respectively. Figure~\ref{rmag-r50} shows that most of the galaxies brighter than $r \sim 19$ mag,
 detected in the UVIT image are members of the cluster \abell. The bright stars ($r \lesssim 20$ mag) on the other hand, are well separated from the galaxies in this magnitude-radius space. 
 We note that the faint stars within $\sim 2$ magnitudes from the completeness limit of the SDSS, might be high redshift or faint galaxies, misclassified as stars by the automated SDSS 
 pipeline. Spectroscopic data or high resolution, deeper imaging of this field can help clarify the status of these sources.


 \section{The \fuv sources detected by UVIT}
 \label{props}
 
  \begin{figure}
 \centering{
 {\includegraphics[scale=0.5]{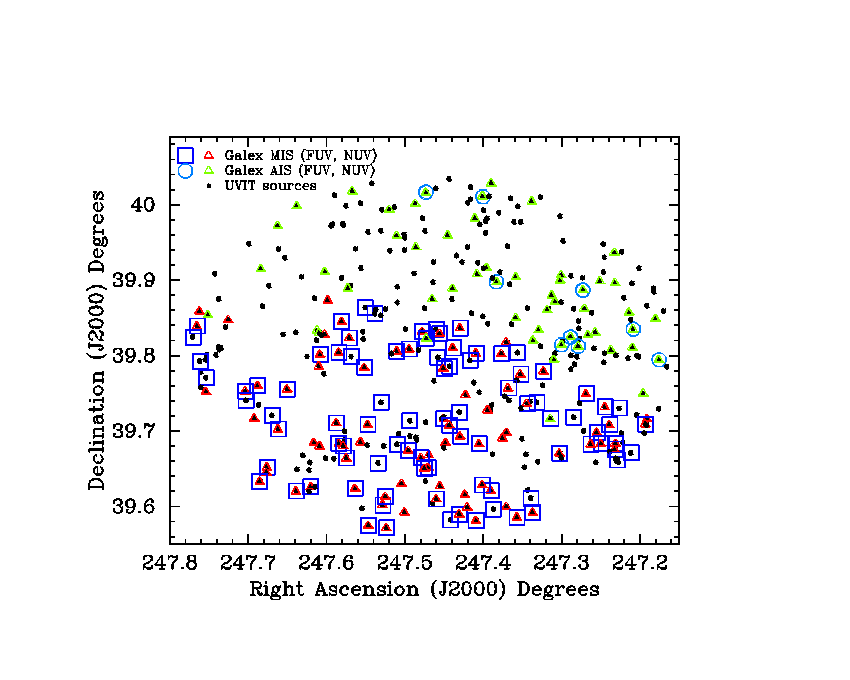}}}
 \caption{All 352 UVIT \fuv sources ({\it black points}) are shown, along with those detected by the \g mission. The {\it blue squares} and {\it red triangles} represent the sources
 detected in the deeper \fuv and \nuv MIS \g image, while the {\it blue circles} and {\it green triangles} depict the \fuv and \nuv sources detected in the AIS \g images, respectively. }
 \label{sky}
 \end{figure}
 
 Prior to UVIT/AstroSat, \abell~has been observed by the \g mission. However, two sets of \g images overlap with different parts of the UVIT field of view. So in order to 
 compare UVIT and \g data, and make use of the \nuv images in this work, we employ all the \g images. The \g mission provides two sets of images with one image each in the far 
 ultraviolet ({\it FUV}) and near ultraviolet ({\it NUV}) bands. One set of images belonging to the all-sky imaging survey (AIS) have an exposure time of 239 seconds ({\it FUV}) and 
 319 seconds ({\it NUV}), respectively, while the medium imaging surveys (MIS) of the \g mission were exposed for 1649 seconds in both the filters. {\sc sextractor} was used to identify 
 sources on each of the \g images individually. We then searched for a \g counterpart within $3^{\prime\prime}$ of each of the UVIT sources independently, prioritising the deeper \g 
 image in regions of overlap. This exercise yielded 99 \fuv and 140 \nuv sources matched between the \g and UVIT images. 
  
 All the UVIT sources, overplotted with symbols representing different \g images are shown in Fig.~\ref{sky}. A little less than the top half of the UVIT field of view overlaps with the shallow 
 AIS images, as a consequence of which many of the UVIT sources were not detected in the \g images. On the other hand, the bottom half of the UVIT field of view is well covered by the \g 
 MIS images, although there still remain a non-negligible number of sources which have been observed in the \fuv band by the UVIT for the first time. This analysis shows the power of 
 UVIT's unprecedented resolution in discovering new \fuv sources.

 \begin{figure}
 \centering{
 {\includegraphics[scale=0.42]{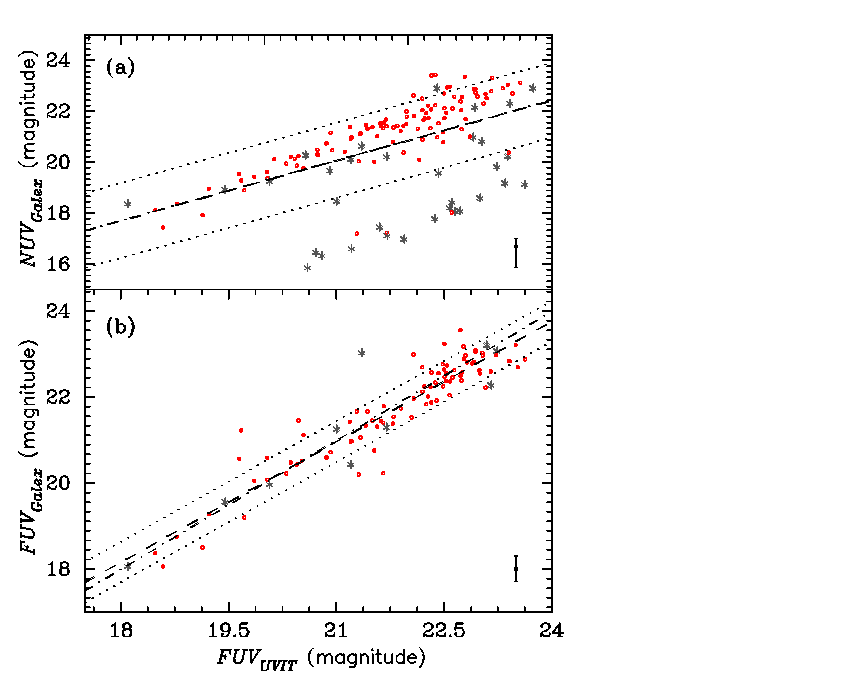}}}
 \caption{The distribution of magnitude of (a) \g \nuv, and (b) \g \fuv sources plotted as a function of the UVIT \fuv magnitude for all the sources detected by both the instruments. 
 All magnitudes are corrected for extinction due to Milkyway. The {\it red points} denote galaxies and {\it grey asterisks} represent the stars, respectively.
 The {\it dot-dashed line} in both panels represents equal magnitudes from both the instruments, while the {\it dashed lines} represent the least square fit to the data, along with $1\sigma$ 
 deviation ({\it dotted lines}) in it. Typical uncertainty along each axis is shown in the bottom right corner of the panels. This figure therefore shows that the \fuv fluxes of the matched 
 sources are consistent between the UVIT BaF$_2$ band and both the wavebands of the \g mission. }
 \label{uvit-galex}
 \end{figure}

 The \g \fuv and \nuv magnitudes of all the matched galaxies and stars are shown as a function of their UVIT \fuv magnitude in Figure~\ref{uvit-galex}. The magnitudes for the sources are 
 well matched, with many of the stars forming a separate sequence in the {\it FUV}$_{UVIT}$-{\it NUV}$_{\it Galex}$ plane. 
 As expected, the scatter at the faint end is higher than the bright end for 
 both the distributions. In fact, the standard deviation of the correlation is clearly dictated by the faint-end scatter, especially in the {\it FUV}$_{UVIT}$-{\it NUV}$_{\it Galex}$ plane. The typical
 uncertainty in the \g \fuv and \nuv filters are $\sim 0.3$ and $0.6$ mag, respectively, which is more than an order of magnitude higher than the uncertainty of $\sim 0.02$ mag found for the 
 UVIT BaF$_2$ magnitudes reported here.

 \begin{figure}
 \centering{
 {\includegraphics[scale=0.5]{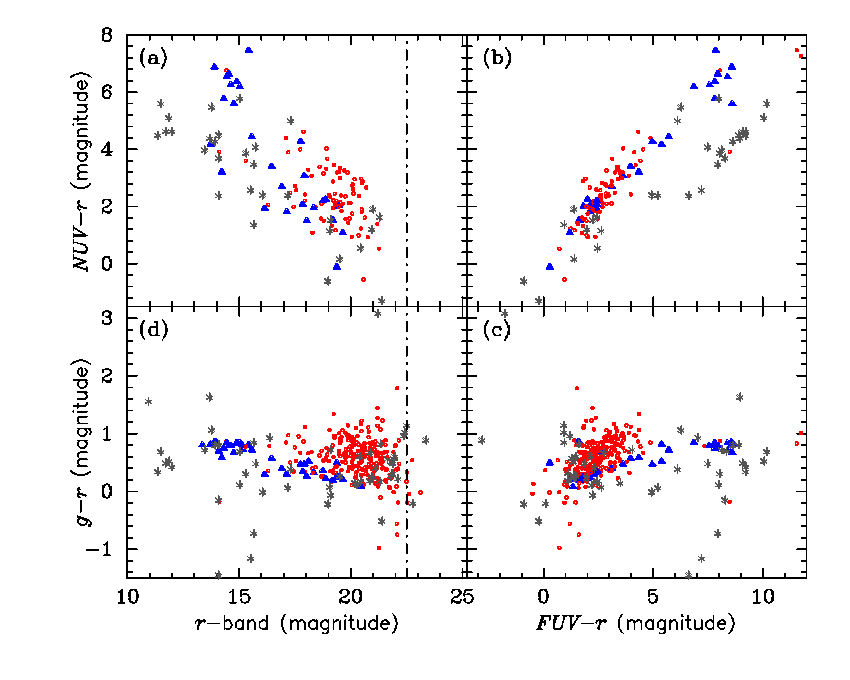}}}
 \caption{The distributions for all UVIT sources along with their optical and \g counterparts in different colour--colour and colour--magnitude planes. Clockwise from top left, we have shown
 (a) $NUV-r$ vs $r$-band magnitude, (b) $NUV-r$ vs $FUV-r$, (c) $FUV-r$ vs $g-r$, and (d) $g-r$ vs $r$-band magnitude plane, respectively. The symbols are same as in Fig.~\ref{rmag-r50}. 
 The $FUV$ magnitudes are from UVIT, while the \nuv magnitudes are from the \g mission. The number of data points in each quadrant will vary 
 as mentioned in Sec.~\ref{data}. While the \abell~cluster galaxies span the entire range of colours and magnitudes available, the other galaxies are relatively fainter
 and bluer in comparison. It is notable that galaxies and stars are not easily distinguishable in these planes, except for one 
 exception. A handful of stars redder than $NUV-r \sim 2$ mag and $FUV-r \sim 5$ mag, seem to form a separate sequence parallel to the galaxies in the {\it FUV-r} vs {\it NUV-r} plane. 
 }
 \label{col-col}
 \end{figure}

 All the stars and galaxies observed in the UVIT image are plotted in planes mapped by various colours and magnitudes in Figure~\ref{col-col}. Note that the different panels comprise different 
 number of data points as per the availability. The \abell~cluster galaxies span the entire range covered by the $r$-band magnitude and the {\it NUV-r} colour (Figure~\ref{col-col} (a)). It is 
 interesting to note that even the stars occupy the entire range of $r$-band magnitude, making it impossible to distinguish them from galaxies. The other galaxies coincide with the cluster
 galaxies, but are on average fainter than their counterparts in the cluster. It is interesting to note that these results are in contrast with a similar analysis performed for the Coma 
 cluster \citep[Abell~1656, $z=0.023$;][]{mahajan22}. In figure~7 of \cite{mahajan22}, the bright galaxies in the Coma cluster clearly separated from the other galaxies. This discrepancy 
 may be due to the fact that: (i) higher number of sources detected in the Coma field improved the statistics, (ii) the brightest Coma galaxies are $M_r \sim 3$ mag brighter than the ones 
 detected in \abell, providing for a better contrast than available in the sample used here, (iii) Coma is a more evolved cluster relative to \abell, (iv) a combination of the above.  

 All galaxies form a rather narrow sequence in the {\it FUV-r} vs {\it NUV-r} colour-colour plane (Figure~\ref{col-col} (b)).
 In this case however, the stars redder than $NUV-r \sim 2$ mag and $FUV-r \sim 5$ mag, form a separate sequence parallel to the galaxies, similar to what was observed for the 
 Coma field \citep{mahajan22}. 

 The {\it FUV-r} vs {\it g-r} colour-colour plane (Figure~\ref{col-col} (c)) shows that the bright red cluster galaxies are somewhat separated from their fainter counterparts, which coincide well 
 with the other galaxies. The stars on the other hand, do not show any trend, and are scattered everywhere in this plane.  
 
 In the optical colour-magnitude plane (Figure~\ref{col-col} (d)) the cluster galaxies form the expected red sequence, while the other galaxies show considerable scatter in colour at fixed magnitude.
 We note that the observed cluster galaxies form a red sequence overlapping the one observed using optical data alone \citep[][see their figure~1 for comparison]{song17}. 
 Also, most of the other galaxies are fainter than the cluster galaxies, such that the brightest galaxies observed in the UVIT field are invariably classified as members of the cluster \abell. 

 
 \begin{figure}
 \centering{
 {\includegraphics[scale=0.5]{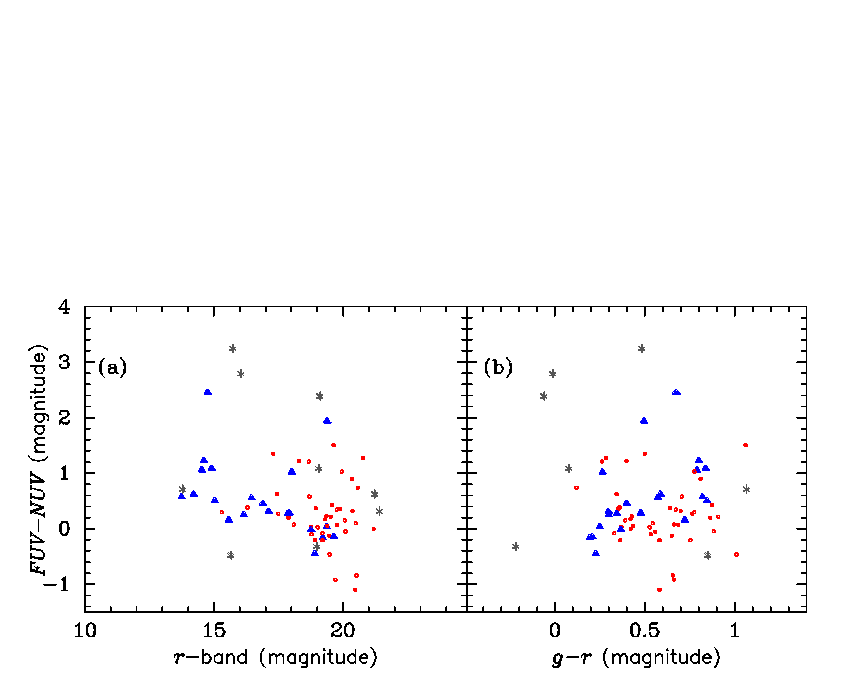}}}
 \caption{The distribution of all the UVIT sources having optical and \g counterparts in the (a) {\it FUV-NUV} vs $r$-band and (b) {\it FUV-NUV} vs {\it g-r} colour-colour plane, 
 respectively. The symbols are same as in Fig.~\ref{rmag-r50}. With the exception of three outliers, the \abell~cluster galaxies form a sequence, such that the fainter galaxies have bluer UV colour, 
 the other galaxies and handful of stars making an appearance in these planes do not show any trend. 
 }
 \label{fnuv}
 \end{figure}

 In Figure~\ref{fnuv} we show the distribution of the {\it FUV-NUV} colour of the cluster galaxies, other galaxies and stars as a function of their $r$-band magnitude and the optical {\it g-r} colour, 
 respectively. It is interesting to note that while most of the cluster galaxies do seem to form a sequence such that the fainter galaxies have bluer UV colour, the other galaxies are very scattered in this
 space. This observation is in contrast to a similar diagnostic for the Coma cluster \citep[see figure 8 of][]{mahajan22}, where the cluster galaxies were found to occupy well defined regions in this 
 plane. This discrepancy is partly due to the lower number density of sources observed in the image used in this work, relative to the Coma field.


 \section{Multi-wavelength observations of distorted candidates in \abell}
 \label{spl}
 
  \begin{table}
 \caption{UVIT sources with LoTSS (144 MHz) radio counterparts.}
 \begin{center}
 \begin{tabular}{ lccccr }     
 \hline
 Galaxy & Right Ascension  & Declination  & Redshift  & 144 MHz flux  & Source of flux \\ 
         &  (J2000)                    & (J2000)      &    ($z$)     &   (mJy)  &     \\
 \hline
UGC 10420 	& 	247.4627	 &	39.766	&   0.0306  &	 $74.67$   &  \cite{mahajan23} \\ 
UGC 10429	&	247.6387	 &	39.831	&   0.0246   &    $26.40$	&  \cite{roberts21}	\\ 
\mcg 		&	247.2091   &    39.835       &  0.0359    &    $20.40$  	&  \cite{roberts21}   \\ 
Z 224-55		&	247.6200 	&	39.626	&   0.0304	   &	 $8.50$	&  \cite{shimwell22}	  \\
Z 224-65		& 	247.7642  & 	39.838 	&   0.0305  &     $6.10$	&  \cite{shimwell22}	  \\
SDSS J163017.01$+$394924.3 & 247.5709 &  39.823   &   0.0701 &  $3.20$   &  \cite{shimwell22}	 \\
\hline
\end{tabular}
 \end{center}
 \label{tab:radio}
 \end{table}

  \begin{figure}
 \centering{
 {\includegraphics[scale=0.49]{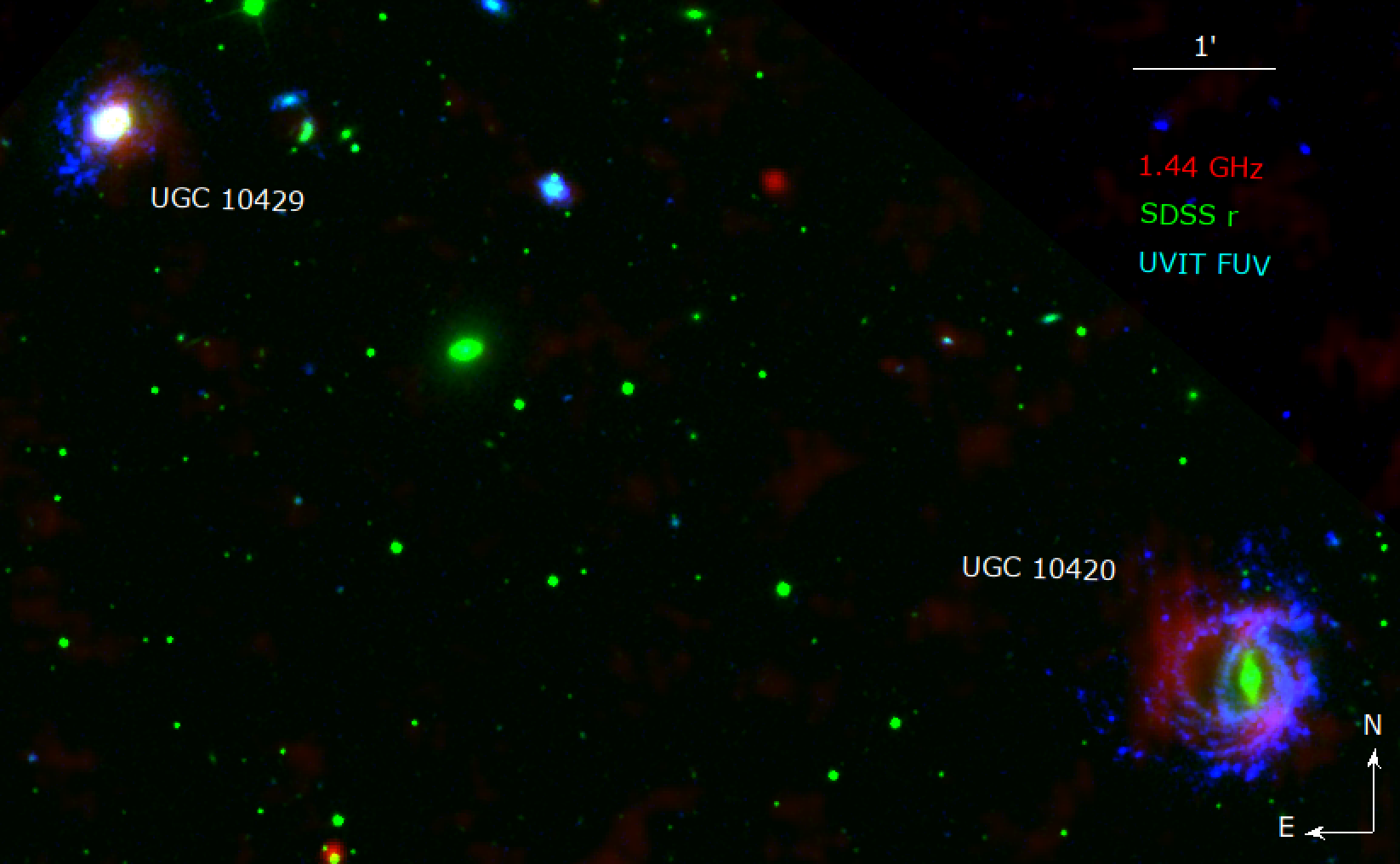}}}
 \caption{A multi-wavelength view of galaxies UGC 10420 and UGC 10429 in \abell. The {\it rgb} colours represent 144 MHz emission {\it (red)},
 SDSS $r$-band {\it (green)} and the UVIT \fuv {\it (blue)}, respectively.  }
 \label{ugc}
 \end{figure}

 \begin{figure*}
 \centering{
 {\includegraphics[scale=0.2]{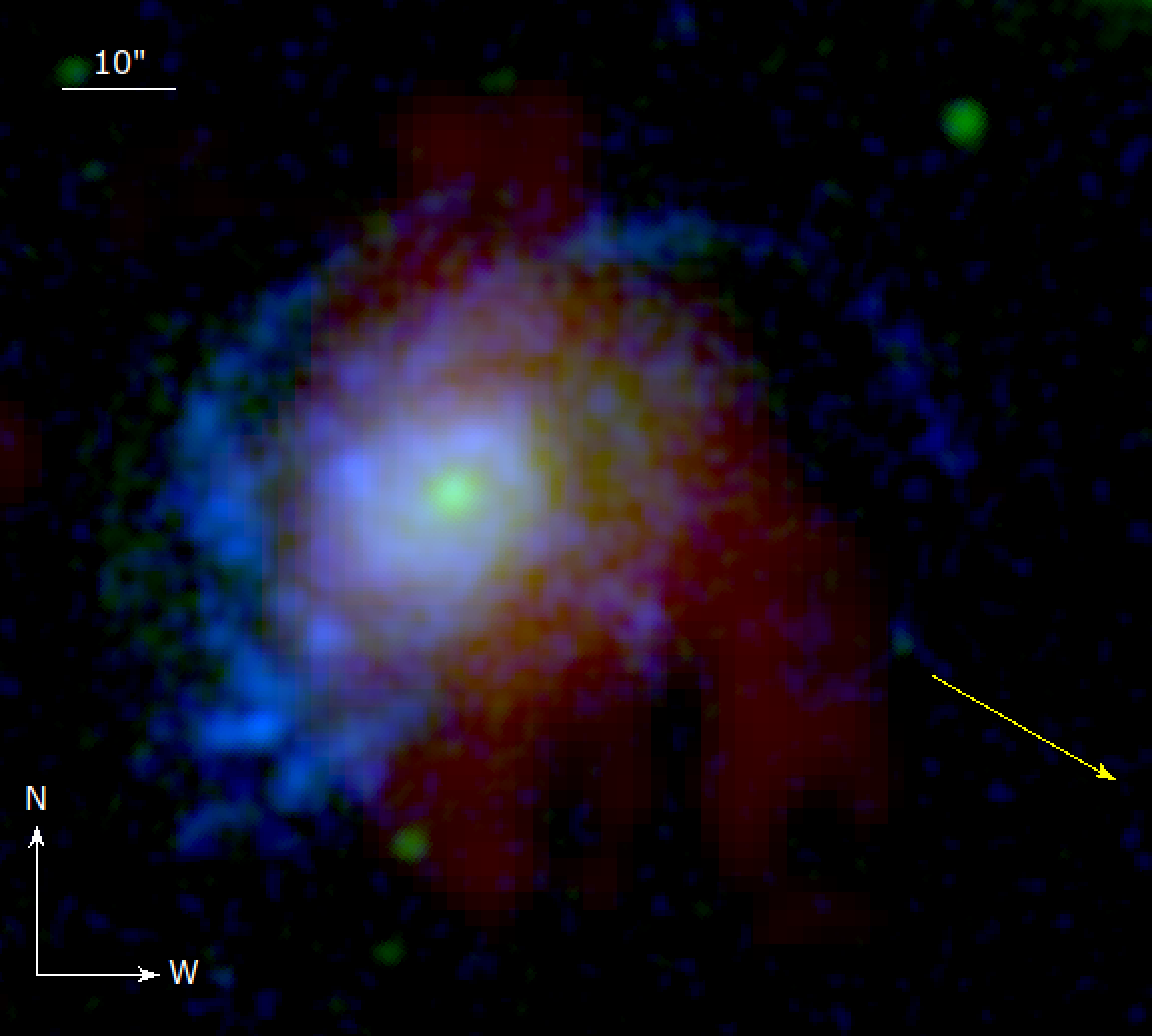}}}
\centering{
 {\includegraphics[scale=0.235]{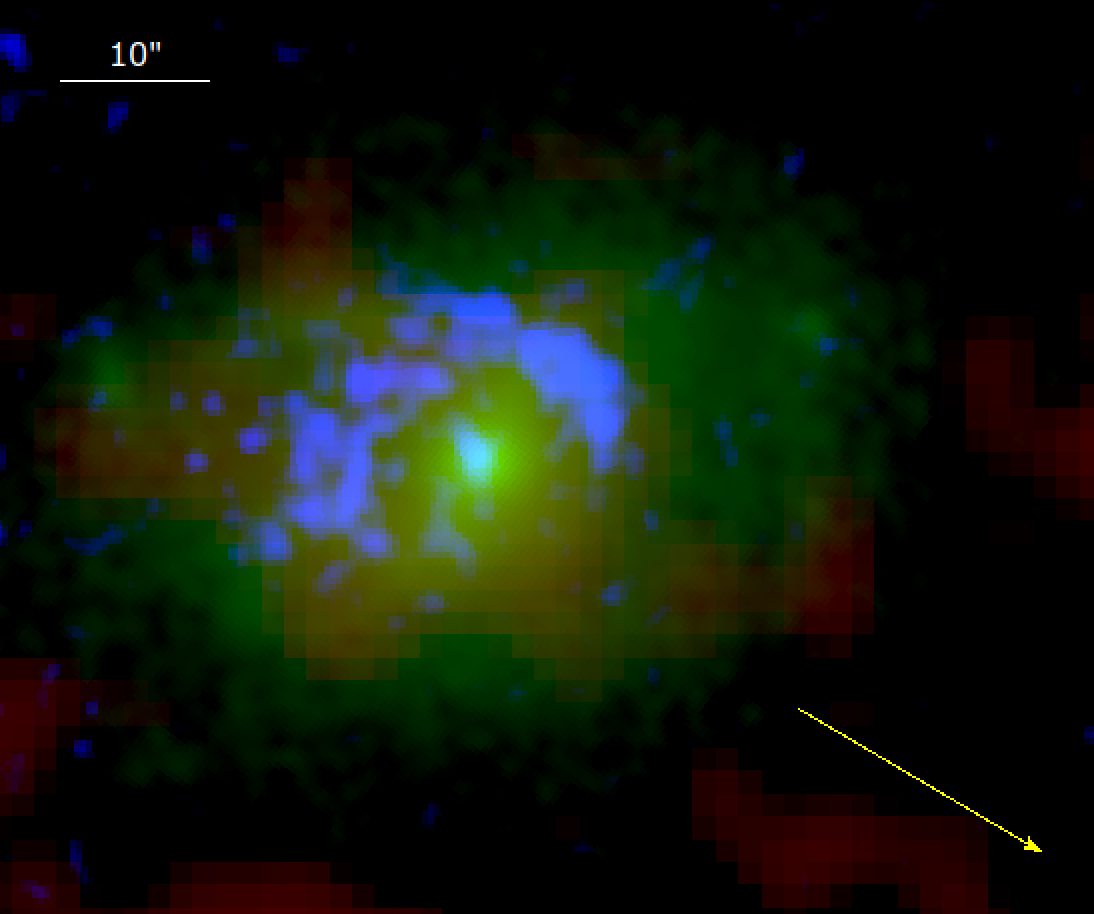}}}
 \centering{
 {\includegraphics[scale=0.28]{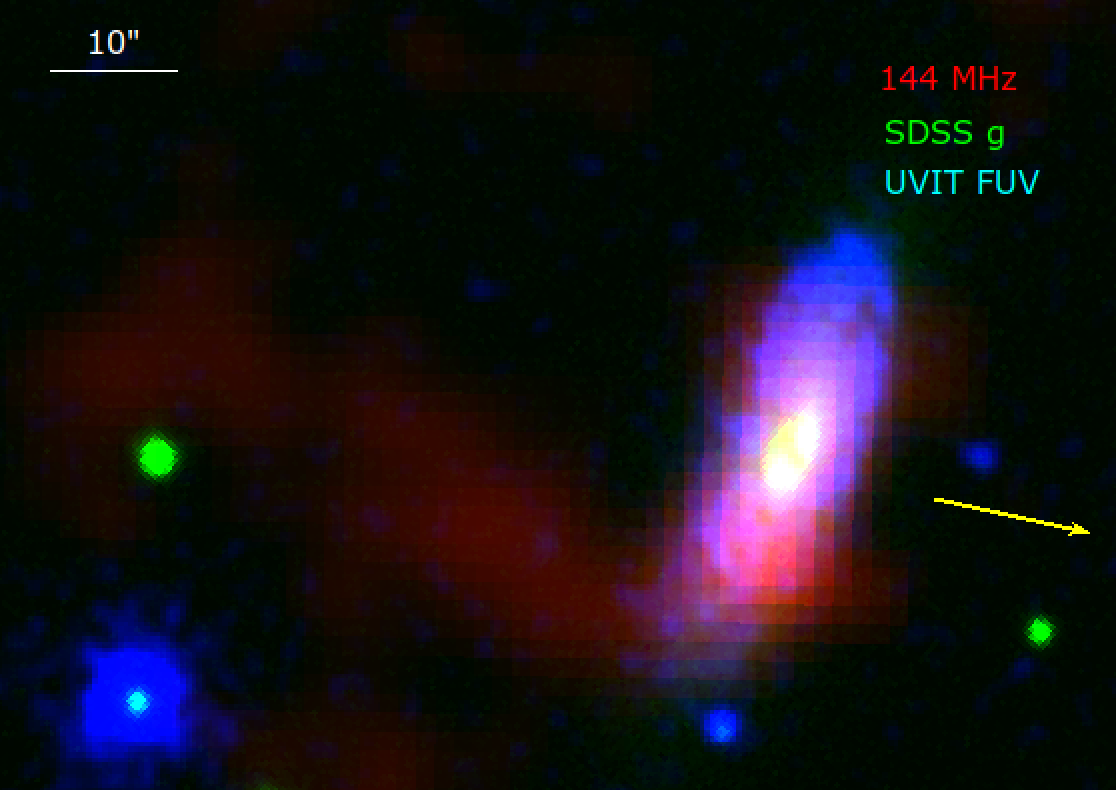}}}
 \caption{Multi-wavelength view of RPS candidate galaxies in \abell. {\it (left to right:)} UGC 10429, Z224-65 and Z224-55, respectively. The {\it yellow} arrow points in the direction 
 of the cluster centre. While the radio tail of Z224-55 points opposite the cluster centre, the \fuv emission in the other two galaxies follow suit, suggesting an outburst of star formation
 as a consequence of interaction with the intra-cluster medium. }
 \label{deformed}
 \end{figure*}

 We searched the literature for existing observations of galaxies in our UVIT image, and found radio counterparts for six galaxies in the LOw-Frequency ARray (LOFAR) Two-metre 
 Sky Survey (LoTSS) data \citep[120-168 MHz;][]{shimwell17, shimwell22}. The positions, redshift and LOFAR (144 MHz) fluxes for these galaxies are listed in Table~\ref{tab:radio}. 
 
 \subsection{UGC~10420 and 10429}
 
 UGC~10420 is a face-on spiral galaxy undergoing ram-pressure stripping in \abell~(Figure~\ref{ugc}). We have presented detailed multi-wavelength analysis of UGC~10420 in 
 \cite{mahajan23}, where we showed that it is not only an RPS candidate, but the only known cluster galaxy with an extended UV disk.  
 
 \cite{roberts21} have recently presented radio data, taken from LoTSS for nearby ($z < 0.05$) 
 jellyfish\footnote{In the context of this paper, the term `jellyfish' is used for galaxies with unambiguous tail(s) of material stripped from their disc. These galaxies are believed to be a
 product of RPS in clusters.} galaxies. 
 One of the cluster galaxies from their sample observed by the UVIT is UGC~10429 (Figures~\ref{ugc} and \ref{deformed}(left)).
 UGC~10429 having $M^* = 10^{10.5} M_{\odot}$ \citep{salim16}, is
 $\sim 2100$ km s$^{-1}$ away from the nearest large galaxy UGC~10420 in redshift space (Figure~\ref{ugc}). The projected distance between UGC~10429 and UGC~10420 is 
 $\sim 347~h^{-1}$ kpc, which diminishes a chance of ongoing interaction between them. These two galaxies however, may have had a close flyby interaction in the past. 

 UGC~10429 is a low-ionization nuclear emission region galaxy \citep[LINER;][]{toba14,zaw19}. Interestingly, while the low-frequency radio emission in UGC~10420 appears behind the 
 galaxy relative to the optical disc \citep[see,][for a detailed discussion]{mahajan23}, in UGC~10429 the radio emission is observed at the edge of the optical disc facing the cluster 
 centre (Figure~\ref{deformed}(left)). On the other hand, the \fuv emission is more intense on the side of the disc away from the cluster centre.      
 
 \begin{figure}
 \centering{
 {\includegraphics[scale=0.39]{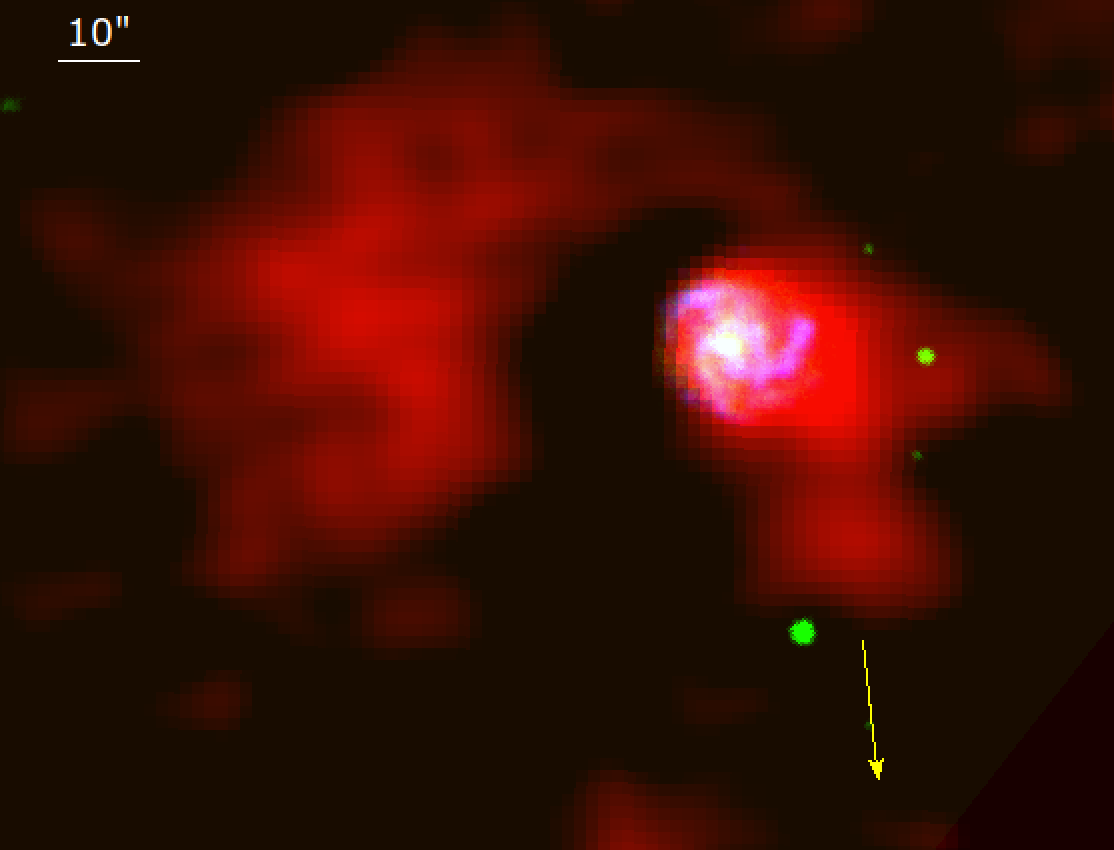}}}
 \centering{
 {\includegraphics[scale=0.65]{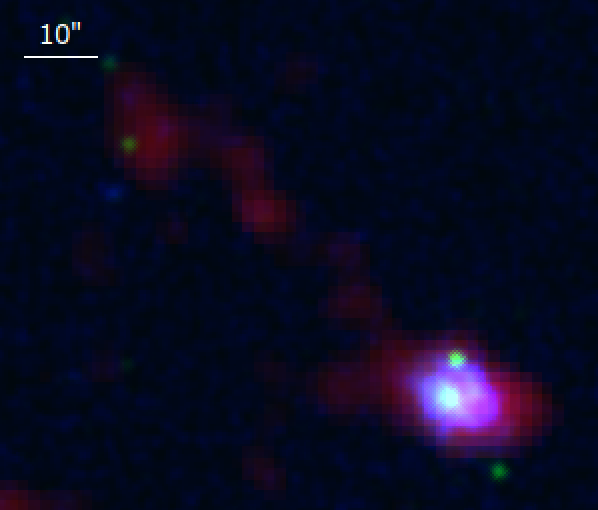}}}
 \caption{{\it left:} The Sd spiral galaxy \mcg, with large-scale radio emission associated with it is classified as a jellyfish galaxy by \cite{roberts21}. {\it right:} Spiral galaxy 
 SDSS J163017.01$+$394924.3, which is not a member of \abell, but could be part of another large-scale structure (see Sec.~\ref{bg-grps}). The {\it rgb} colour scheme for both images 
 is the same as in Figure~\ref{ugc}. }
 \label{spiral}
 \end{figure}

  \subsection{Z224-65}
  
 Z224-65 (Figure~\ref{deformed}(centre)) is a barred Seyfert type II galaxy \citep{toba14}. \cite{lee15} have classified Z224-65 as a mid-infrared (MIR) green 
 valley\footnote{Galaxies lying between the red sequence and blue cloud in a colour-magnitude diagram are popularly known as `green valley' galaxies in the literature. \cite{lee15} 
 classified MIR green valley galaxies using a $\nu L_{\nu}/L_{\odot}~(12 \mu{\rm m})$ vs $[3.4]-[12]$ WISE colour (see their fig.~3).  }
 galaxy based on its Wide-field Infrared Survey Explorer 
 (WISE) data. These authors find that most of the MIR green valley galaxies belong to the optical red sequence. Their analysis suggests that the morphological transition of galaxies 
 from early to late type occurs in the MIR green valley phase. \cite{lee15} also find that the galaxies entering the MIR green valley phase have already been quenched. 
 
 The UVIT \fuv emission from Z224-65 is mostly constrained in an arch around the central bulge, in a direction opposite to the cluster centre. The low-frequency radio emission shows 
 no preference, and is scattered around the optical disc. It would be interesting to obtain a high-frequency radio image of Z224-65 to throw more light on the interaction 
 between the intra-cluster and the interstellar mediums, and confirm if the \fuv emission is linked to the turbulence caused by an environment related mechanism. 
 
  \subsection{Z224-55}
 
 Z224-55 ($z=0.030$; Figure~\ref{deformed}(right)) is classified as a LINER galaxy \citep{toba14} in \abell. \cite{zaw19} classified this galaxy as a Type I AGN based on the broad 
 EW(H$\alpha$) and other emission lines present in the spectra. Z224-55 shows an $80\arcsec$ long low-frequency radio tail in the direction opposite to the cluster centre. These 
 multi-wavelength data suggests that Z224-55 may be undergoing ram-pressure stripping under the influence of the cluster's potential well, at a distance of $\sim 0.8 h^{-1}$ Mpc (almost 
 $0.5R_{200}$) from the cluster centre. 
 
 \subsection{\mcg}
  The second galaxy in common with the sample of \cite{roberts21}, \mcg, is an Sd-type spiral galaxy with large-scale radio emission associated with it (Figure~\ref{spiral}). 
 This galaxy with $M^* = 10^{10.7} M_{\odot}$, and SFR $\sim 3.16~M_{\odot}{\rm yr^{-1}}$ \citep{wang17} is also classified as a jellyfish galaxy \citep{roberts21}. The SDSS spectra 
 for \mcg~reveals large  equivalent width, EW(H$\alpha$), but negligible emission in [OIII], 
 suggesting that this galaxy is not an optical AGN. We also did not find any X-ray emission associated with this object. Therefore, the low-frequency radio emission seems to be 
 associated with the ongoing star formation in the galaxy: either feeding it, or a consequence of a stellar feedback from massive stars, or a combination of both.     
 
 \subsection{SDSS J163017.01$+$394924.3} 
  At redshift $z = 0.0701$, SDSS J163017.01$+$394924.3 is not be a member of \abell. The bright spot seen northward of the core in the optical image of this spiral galaxy is a 
  foreground star with no \fuv emission associated with it (Figure~\ref{spiral}).   
  The low frequency radio emission outside the optical disk is faint, but assumes a tail-like morphology extending in the north-east direction. It is interesting to note that this galaxy will be close 
  in redshift space ($\sim 2000$ km s$^{-1}$) to the large-scale structure which may be present in this field of view at a higher redshift (see Sec.~\ref{bg-grps} and Fig.~\ref{z-hist} below).

  To summarise this section, the data suggests that all five member galaxies are undergoing transformation lead by environment-driven processes in the cluster \abell. 
  SDSS J163017.01$+$394924.3 also shows a low-frequency radio tail and could be a RPS candidate in a higher redshift group. The fact that all the \abell~members 
 lie within the virial radius \citep[$r_{vir} = 1.6 h^{-1}$ Mpc;][]{rines02}, further supports our speculations. Amongst these, Z224-44 $\sim 0.72 h^{-1}$ Mpc from the centre of the cluster
 is the closest, while Z224-65 at $\sim 1.26 h^{-1}$ Mpc is the farthest, respectively.
 
 \section{Star formation rates of FUV detected galaxies}
 
 \begin{figure}
 \centering{
 {\includegraphics[scale=0.3]{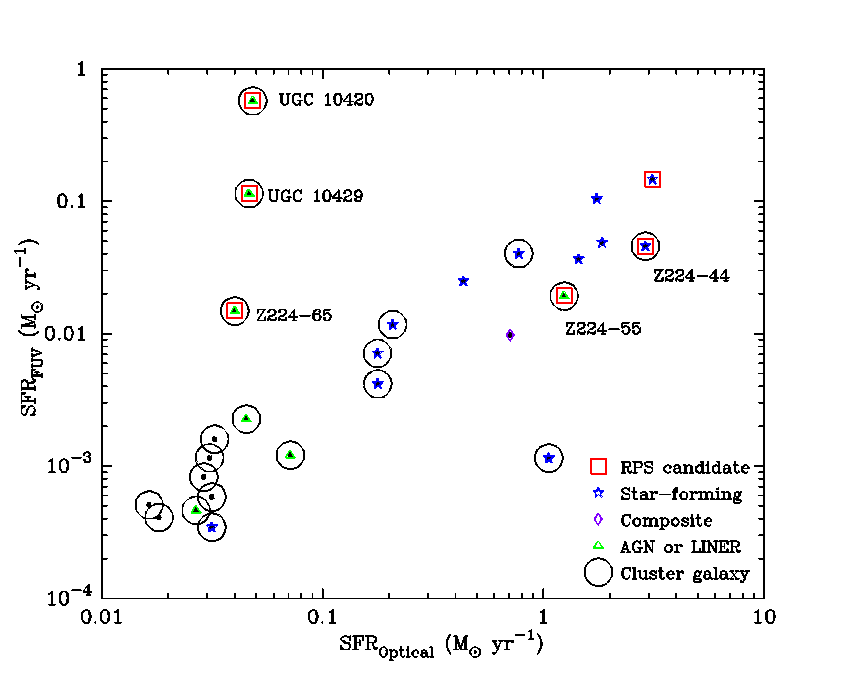}}}
 \caption{SFR of galaxies estimated using \fuv data plotted as a function of the SFR estimated using the SDSS optical data. 20 out of 28 galaxies plotted here are 
 members of the cluster \abell~{\it (black circles)}. The RPS candidate galaxies are marked by {\it red squares}, while the BPT classification is represented as: AGN or LINER {\it (green triangles)}, 
 star-forming {\it (blue stars)} and composite {\it (purple diamond)}, respectively. }
 \label{sfr}
 \end{figure}

 We find 28 galaxies in common with the UVIT data presented here and the MPA-JHU star formation rate (SFR) catalogues\footnote{The MPA-JHU catalogues are produced by a group of 
 researchers currently or formerly at the Max Planck Institute for Astrophysics and thr John Hopkins University. https://wwwmpa.mpa-garching.mpg.de/SDSS/}. Of these, 20 galaxies are
 members of the cluster \abell. Figure~\ref{sfr} shows 27 of these galaxies (one galaxy at [$10^{-3.52}, 10^{-6.89}$] is omitted for clarity), including six RPS candidate galaxies discussed above marked explicitly. 
 The \fuv SFR is estimated using the calibration provided by \cite{iglesias06}. We have not changed the formula to compensate for the UVIT wavebands instead of \g used by them because
 both the missions have \fuv broadbands similar to each other. We have also not applied any extinction correction to the \fuv fluxes other than that mentioned in Sec.~\ref{uvdata}, because 
 the required far infrared data are not available for these galaxies \citep[see,][for a detailed discussion on multi-wavelength SFRs]{mahajan19}. 
 
 As expected, the attenuated SFR$_{\fuv}$ underestimates the SFR for all galaxies by $\sim 2$ orders of magnitude, except three RPS candidates in \abell, namely UGC~10420, UGC~10429
 and Z224-65. On the other hand, Z224-55, Z224-44 and the non-member RPS candidate SDSS J163017.01$+$394924.3 seem to follow the sequence formed by the other galaxies in this 
 SFR$_{optical}$ vs SFR$_{\fuv}$ plane. This analysis therefore suggests that environment-related processes such as RPS may enhance star formation in some of the cluster galaxies.




\section{Background group(s) in the field of view}
\label{bg-grps}

 \begin{figure}
 {\includegraphics[scale=0.25]{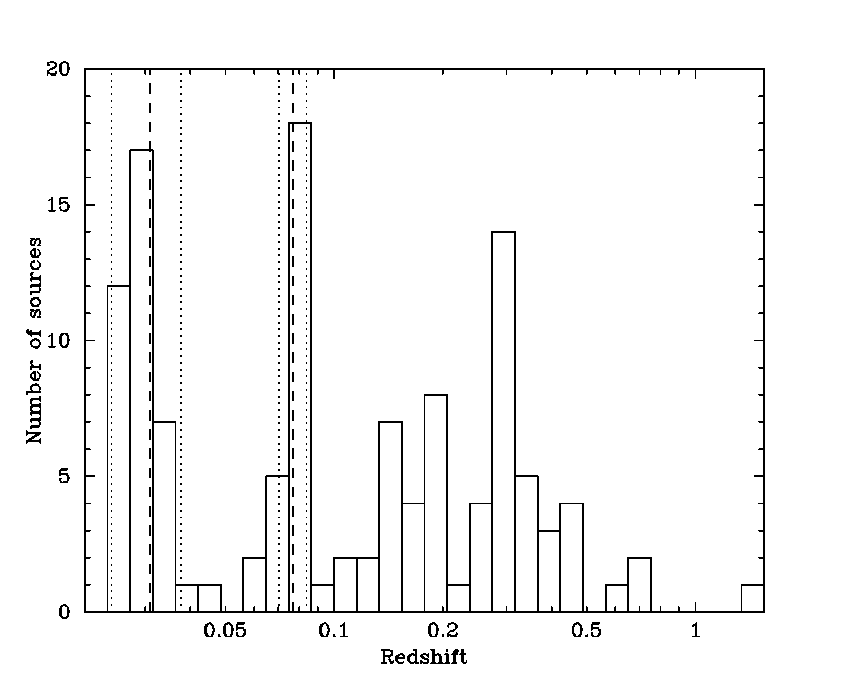}}
  {\includegraphics[scale=0.25]{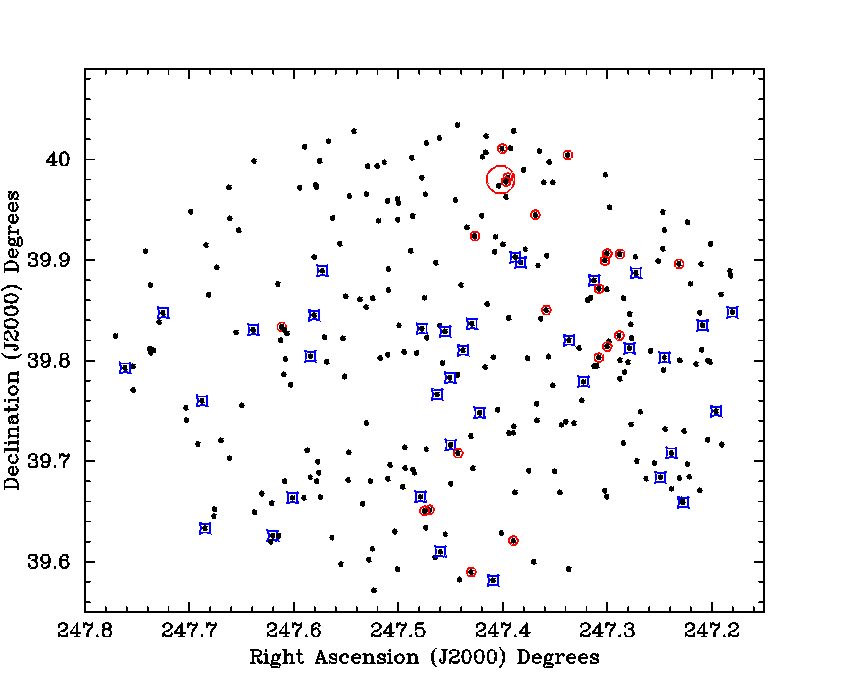}}
 \caption{(a) The redshift distribution of sources with spectroscopic redshift $\ge 0.02$ in the field of view. The first set of lines represent the redshift slice for the cluster \abell 
 ($z_{A2199} \pm 3\sigma_v$). The compact group SDSSCGA~00777 ($z=0.0769$), on the other hand is likely an embedded substructure in a larger cluster or group (see text for 
 discussion). There are 22 galaxies in a redshift slice of $\pm 2000$ km s$^{-1}$ {\it (dotted lines)} around this group. Another prominent peak in the redshift distribution is also observed 
 $\sim 0.28$, although no documented groups or clusters were identified at this position. (b) Position of all the observed galaxies in the UVIT field of view for \abell. The galaxies 
 highlighted as {\it blue squares} are cluster members, while the ones marked by {\it red circles} represent the galaxies within $\pm 2000$ km s$^{-1}$ of the compact group 
 SDSSCGA~00777 {\it (large red circle)}.  }
 \label{z-hist}
 \end{figure}

 \begin{figure}
 \centering{
 \frame{\includegraphics[scale=0.28]{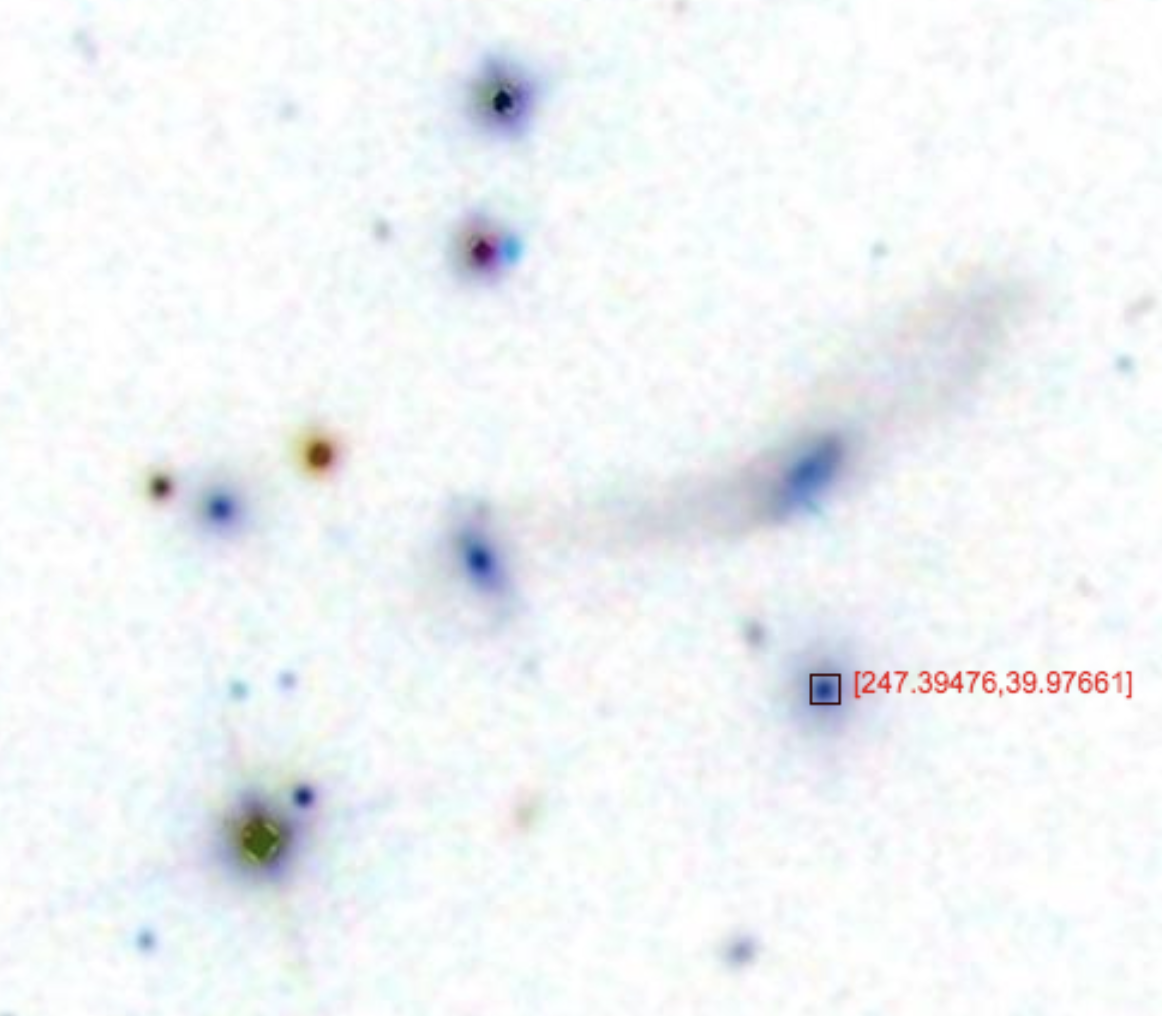}}}
 \caption{The SDSS {\it u,g,r,i,z} image of the compact group SDSSCGA~00777. The coordinates for LEDA~2158671, the only galaxy with redshift, are marked explicitly. While this galaxy
 is not seen by the UVIT, very faint emission from the other two interacting galaxies can be observed in the UVIT image. }
 \label{groupgals}
 \end{figure}

 We noticed several interesting galaxies further away in redshift space from \abell. This observation led us to explore the chance alignment of background structures behind \abell, in this 
 field of view. Figure~\ref{z-hist}(a) shows the distribution of redshift for 122 sources with $z \ge 0.02$. The redshift distribution evidently peaks at $z \sim 0.03$, for \abell, but shows 
 two other peaks at redshift $\sim 0.077$ and 0.260, respectively. 
 
 Further investigation revealed the presence of a photometrically-selected compact group SDSSCGA~00777 
 \citep[$\alpha: 247.402^{\circ}, \delta: 39.980^{\circ}, z = 0.0769$;][]{mcconnachie09}, having four members in the UVIT field of view (Figure~\ref{groupgals}). The group redshift is assigned based on the 
 spectroscopic redshift of one of the four galaxies, LEDA~2158671 (which is not detected by UVIT). It is noticed, however, that a set of 22 UVIT sources with spectroscopic redshifts 
 were observed within a redshift slice of $\pm 2000$ km s$^{-1}$ around this compact group. The position of these galaxies, along with the centre of SDSSCGA~00777 group are shown in
 Figure~\ref{z-hist}(b). It is likely that these galaxies and group are embedded within another large-scale structure behind \abell~along the line of sight of this field.
 The fact that most of these galaxies are confined to the right hand side of the image, of which 16/22 sources are found only within the top right quadrant of the UVIT field of view, 
 further adds to our speculation.
 
 We could not find any documented group or cluster corresponding to the second peak at $z \sim 0.260$. It is, however, worth mentioning that seven UVIT galaxies are found within a 
 redshift slice of $\pm 2000$ km s$^{-1}$ around $z = 0.2794$. Based on the distribution of redshifts, it is also possible that a large-scale filament extending in the redshift space 
 with its axis close to the line of sight is observed projected in this field of view. Further investigation along these lines are however, out of the scope of this work.


 \section{Summary and results}
 
 In this paper, we have studied an {\it AstroSat}/UVIT field located in the X-ray bright cluster \abell~($z = 0.031$). We have used optical photometric data from the SDSS,
 spectroscopic data compiled by \cite{song17} and low-frequency radio data from the LoTSS (DR2) to study properties of objects detected by the UVIT in the BaF$_2$ \fuv
 filter ($\lambda_{mean} = 1541~\AA;~\Delta \lambda = 380~\AA$). A complete catalogue of 352 sources detected in this UVIT field, along with the optical data sourced from 
 the literature is presented. We find that 35\% of the UVIT sources have spectroscopic redshifts, based on which they have been classified into member and non-member galaxies.  
 The source type identification for the rest of the data are based on the classification provided by the SDSS pipeline, which determines if the source detected in the 5-band SDSS image is 
 resolved (extended) or unresolved (point-like). The pipeline also folds in the information about colours to distinguish stars from quasars. It is noteworthy that some of the non-member 
 galaxies may be cluster galaxies, and, some of the stars identified by the SDSS photometric pipeline may be faint galaxies. However, deeper spectroscopic and better imaging data are required 
 to confirm the status of these objects. 
 
 We summarise the main results from our analysis below.     
 \begin{itemize}
 
 \item
 Galaxies other than the members of \abell~are on average fainter than the cluster galaxies. However, all stars and galaxies occupy the same range of colour and magnitude in the 
 {\it r} vs {\it NUV-r} plane, making them indistinguishable.
 \item
 Some of the  bright stars form a distinct sequence separated from the galaxies by $\sim 1.5$ mag in the {\it FUV-r} vs {\it NUV-r} colour-colour plane.
 \item 
 Besides UGC~10420 discussed in \cite{mahajan23}, five new candidates have been identified in the newly acquired UVIT data as galaxies undergoing environment-related 
 gas stripping. Our hypothesis is based on the asymmetric morphology of these galaxies in the UVIT {\it FUV} and low-frequency radio images, and the fact that all member galaxies lie within 
 the virial radius of \abell, where they are more likely to experience the wrath of the cluster environment. Two of these galaxies are LINERS (UGC~10429 and Z224-55), while one 
 (Z224-65) is a known Seyfert galaxy. SDSS J163017.01$+$394924.3, which is also an RPS candidate does not belong to \abell, but could be a member of the large-scale structure behind
 \abell~in this field of view. 
 \item
 A comparison between optical and \fuv SFR of UVIT detected galaxies shows that the attenuated SFR$_{\fuv}$ for most of the galaxies are underestimated relative to the optical 
 estimate by $\sim 2$ orders of magnitude. Moreover, while SFR$_{\fuv}$ of three RPS candidate galaxies in \abell~are comparable to their SFR$_{optical}$, the remaining three follow the 
 same trend as other galaxies in the SFR$_{\fuv}$ vs SFR$_{optical}$ plane. 
 \item
 Our analysis suggests the presence of large-scale structures other than \abell~in the UVIT field of view. One of these is a larger cluster or group in which the compact group 
 SDSSCGA~00777 is embedded ($z\sim 0.077$). There may also be another system present around $z \sim0.260$, but no corresponding documented structure could be confirmed.
 \end{itemize}
 
 Further spectroscopic investigation and deeper multi-wavelength imagery can throw more light on our results. In particular, it would be very interesting to explore a 
 wider field covering this region to confirm the presence of a group or cluster around $z \sim 0.077$ and $0.260$. 
 
\acknowledgments

 S. Mahajan was funded by the SERB Research Scientist (SRS) award (SB-SRS/2020-21/56/PS), Department of Science and Technology (DST), Government of India. Prof. K. P. Singh thanks the Indian National Science Academy for support under the INSA Senior Scientist Programme. 
  
 This publication uses data from the {\it AstroSat} mission of the Indian Space Research Organisation (ISRO), archived at the Indian Space
 Science Data Centre (ISSDC). UVIT project is a result of collaboration between IIA (Bengaluru), IUCAA (Pune), TIFR (Mumbai), several  
 centres of ISRO, and the Canadian Space Agency (CSA). 
 This research has made use of the NASA/IPAC Infrared Science Archive, which is funded by the National Aeronautics and Space Administration and 
 operated by the California Institute of Technology.  The {\sc topcat} software \citep{taylor05} was used for some of the analysis presented in this paper.  
 We acknowledge the use of LOFAR data in accordance with the credits given on \url{https://lofar-surveys.org/credits.html}, and SDSS data given on 
 \url{https://www.sdss.org/collaboration/citing-sdss/}. 


\bibliographystyle{JHEP}
\bibliography{a2199}

\end{document}